\documentclass[aps,twocolumn,superscriptaddress,showpacs,pre,nofootinbib,floatfix,usenatbib]{revtex4}
\usepackage{graphics}
\usepackage{graphicx}
\usepackage{epsfig}
\usepackage{amsmath}
\usepackage{amssymb}

\def\lsim{\mathrel{\rlap{\lower4pt\hbox{\hskip1pt$\sim$}}
    \raise1pt\hbox{$<$}}}         
\def\gsim{\mathrel{\rlap{\lower4pt\hbox{\hskip1pt$\sim$}}
    \raise1pt\hbox{$>$}}}         

\def\lapprox{\mathrel{\mathop  {\hbox{\lower0.5ex\hbox{$\sim$}
\kern-1.1em\lower-0.7ex\hbox{$<$}}}}}
\def\gapprox{\mathrel{\mathop  {\hbox{\lower0.5ex\hbox{$\sim$}
\kern-1.1em\lower-0.7ex\hbox{$>$}}}}}

\begin{document}

\title{
Probing galactic cosmic ray distribution with TeV gamma-ray sky}
\author{M. Cataldo}
\affiliation{L'Aquila University, Physics and Chemistry Department, 67100 L'Aquila, Italy}
\affiliation{INFN, Laboratori Nazionali del Gran Sasso, 67100 Assergi (AQ),  Italy}
\author{G. Pagliaroli}
\affiliation{Gran Sasso Science Institute, 67100 L'Aquila, Italy}
\affiliation{INFN, Laboratori Nazionali del Gran Sasso, 67100 Assergi (AQ),  Italy}
\email{giulia.pagliaroli@gssi.it}
\author{V. Vecchiotti}
\affiliation{Gran Sasso Science Institute, 67100 L'Aquila, Italy}
\affiliation{INFN, Laboratori Nazionali del Gran Sasso, 67100 Assergi (AQ),  Italy}
\author{F.L. Villante}
\affiliation{L'Aquila University, Physics and Chemistry Department, 67100 L'Aquila, Italy}
\affiliation{INFN, Laboratori Nazionali del Gran Sasso, 67100 Assergi (AQ),  Italy}
\email{francesco.villante@lngs.infn.it}

\begin{abstract}
The distribution of cosmic rays in the Galaxy at energies above few TeVs is still uncertain and this affects the expectations for the diffuse gamma flux  produced by hadronic interactions of cosmic rays with the interstellar gas.
We show that the TeV gamma-ray sky can provide interesting constraints.
Namely, we compare the flux from the galactic plane measured by Argo-YBJ, HESS, HAWC and Milagro with the expected flux due to diffuse emission and point-like and extended sources observed by HESS showing that experimental data can already discriminate among different hyphoteses for cosmic ray distribution. The constraints can be strengthened if the contribution of sources not resolved by HESS is taken into account. 
\end{abstract}

\maketitle

\section{Introduction}
\label{Sec:Introduction}

Cosmic rays (CR) that propagate in different regions of the Galaxy interact with the ambient gas through hadronic processes and produce a diffuse flux of high energy (HE) gamma and neutrinos. These particles propagate to Earth along straight lines providing us with information on the CR space and energy distributions and, thus, in turn, on the CR transport in the galactic magnetic field. 

The diffuse gamma-ray flux has been measured by several experiments, such e.g. \cite{Kraushaar:1972,Kniffen:1973,Mayer-Hasselwander:1982,Hunter:1997}. The Fermi-LAT instrument has recently reported a detailed picture of the entire gamma-ray sky in the multi-GeV domain \cite{Ackermann:2012,Acero:2016} that can be compared with expectations obtained by modelling CR propagation in the Galaxy. Recent analyses pointed out that the spectral distribution of CR may depend on the galactocentric distance. The possibility of a progressive large-scale hardening towards the inner Galaxy was considered in \cite{Gaggero:2015} and confirmed by two different analyses \cite{Acero:2016,Yang:2016} of the Fermi-LAT data. A very recent work \cite{Aharonian:2018rob} argues that CR hardening is not a global large-scale effect but may be the result of local enhancements of CR density in giant molecular clouds.

In any case, the hypothesis that CR spectral index in different regions of the Galaxy is lower (in modulus) than the local value could clearly have a dramatic effect at TeV energies and above. It would enhance the expected HE emission with respect to standard expectations and, moreover, it would produce a different angular distribution of HE photons from the Galactic plane with respect to lower energy observations. 
It is thus important to compare predictions obtained in this assumption with present observational constraints. 

The presence of spectral hardening can be the signature of non-standard CR propagation, as for example CR self-confinement via streaming instability \cite{Recchia:2016} or anisotropic CR transport due to the change of the magnetic field components toward the Galactic center  \cite{Cerri:2017joy}. While the second scenario is expected to hold at all energies, the first one is less relevant with increasing the energy, so that the analysis of high energy gamma data can provide an handle to distinguish between the two mechanisms. A recent work \cite{Pothast:2018} suggests that a large-scale hardening is present up to $\sim 200$ GeV and resilient to parameters adopted in the analysis of the Fermi-LAT data. 

The goal of this paper is to test expectations for the galactic HE gamma-ray emission against existing observational data in the TeV domain. We consider, in particular, the observations provided by Argo-YBJ \cite{Bartoli:2015}, HESS \cite{Abramowski:2014}, Milagro \cite{Abdo:2008} and the preliminary results by HAWC \cite{HAWCdata}. We calculate the diffuse HE gamma flux by using the phenomenological approach introduced in \cite{Pagliaroli:2016, Villante:2017, Pagliaroli:2017} that allows us to implement different assumptions for the CR space and energy distribution in a simple and direct way, including the possibility of a position dependent CR spectral index. 
A similar approach was recently used by \cite{Lipari:2018}. 
We improve with respect to previous calculations \cite{Pagliaroli:2017, Lipari:2018} by considering more realistic descriptions of CR distribution and by discussing the contribution of pointlike and extended gamma-ray sources observed by HESS~\cite{HESS:2018} when comparing to observational data. 
 

The plan of this paper is as it follows. In the first section, we
describe how we calculate the diffuse gamma flux for the different assumptions on the CR spatial and spectral distributions.
In Sec.\ref{sec2} we discuss the contribution due to resolved sources.
In Sec.\ref{sec3} we compare our predictions for the total gamma flux with observational data provided by Argo-YBJ, HESS, HAWC, and Milagro.
Results are discussed in details in Sec.\ref{Sec:Discussion}. Finally we summarize our work in the last section.

\begin{figure*}[!t]
\begin{center}
\includegraphics[width=8cm]{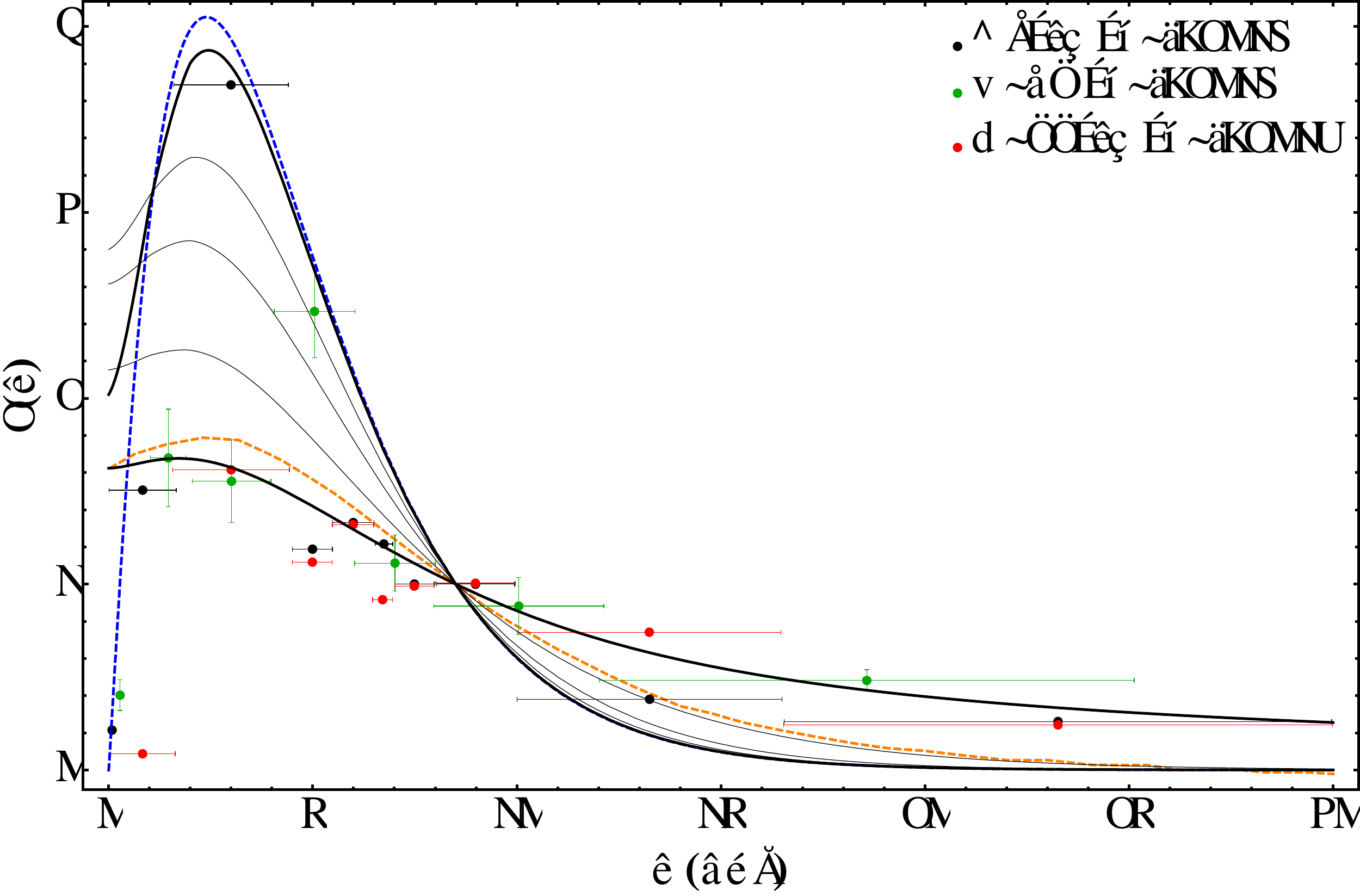}
\includegraphics[width=8.15cm]{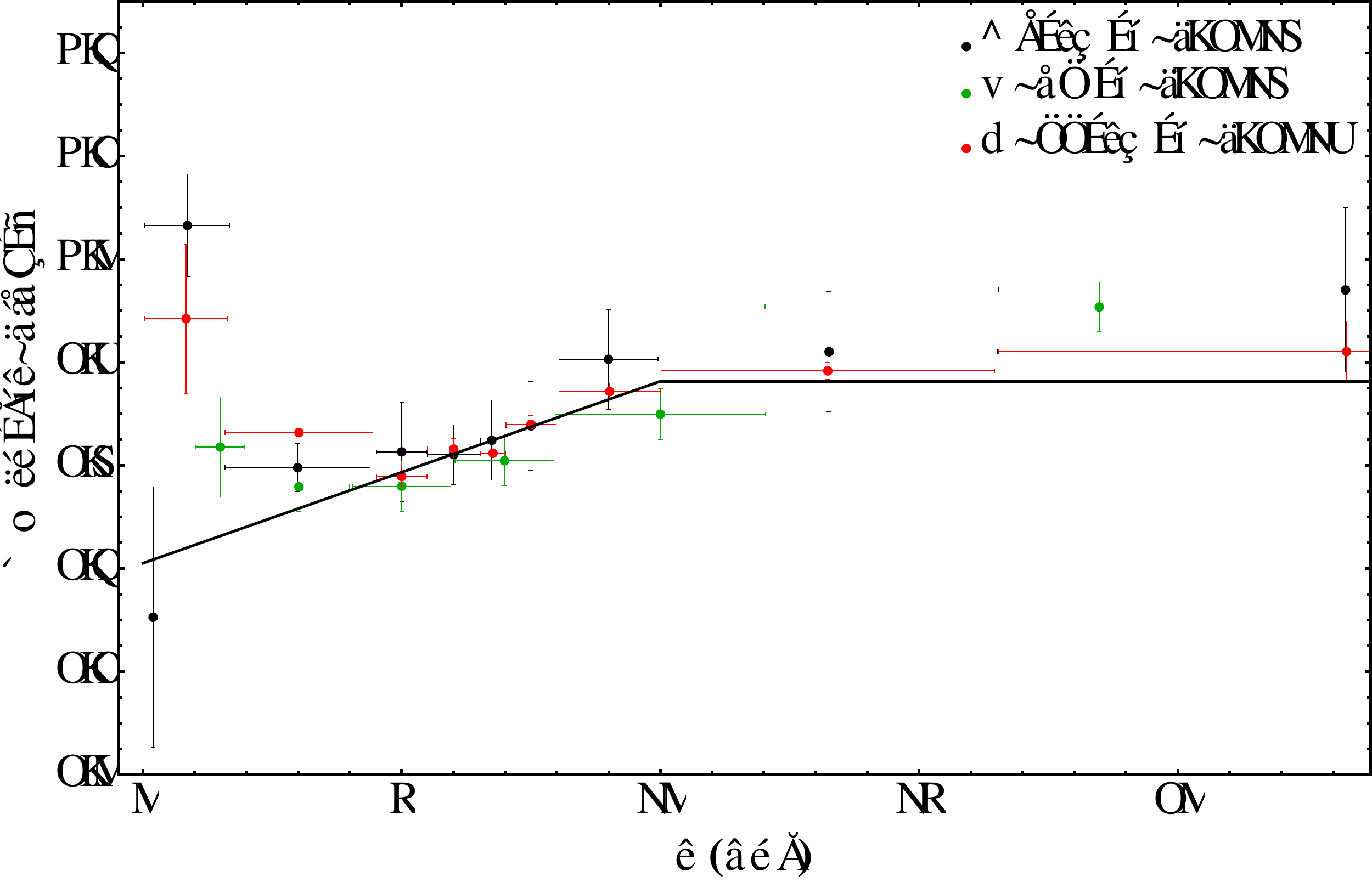}
\caption{\small\em 
\texttt{\rm Left Panel}: 
The black lines show the functions $g({\bf r})$ obtained from Eq.(\ref{Eq:g_funct}) for different smearing radii $R$. Going from top to bottom at $r\simeq 2 kpc$, the different lines correspond to $R=1,3,5,10,\infty$ kpc, respectively. The data points show the CR density at $E \simeq 20\,{\rm GeV}$ \cite{Acero:2016} (normalized to one at the Sun position) and the $\gamma$-ray emissivity above $E_\gamma=1\,{\rm GeV}$ \cite{Yang:2016,Pothast:2018}, as a function of the galactocentric distance obtained from Fermi-LAT data. 
\texttt{\rm Right Panel}:
The black line gives the spectral index of CR at $E_{\rm CR}= 20\,{\rm GeV}$ adopted in this calculation. The data points show the CR spectral index obtained as a function of the galactocentric distance from Fermi-LAT data in \cite{Acero:2016,Yang:2016,Pothast:2018}.  
\label{fig:CR_dist}}
\end{center}
\end{figure*}

\section{The diffuse gamma flux}
\label{sec1}

The diffuse HE gamma flux is calculated according to \cite{Pagliaroli:2016} in the range $E_\gamma \le 20\,$TeV where gamma ray absorption in the Galactic radiation field is expected to be negligible. The main differences with respect to our previous calculation concern the CR distribution in the Galaxy. Following \cite{Pagliaroli:2016}, the CR nucleon flux is given as a function of position and nucleon energy by: 
\begin{equation}
\varphi_{\rm CR}(E,{\bf r}) = \varphi_{\rm CR,\odot}(E)\,g({\bf r})\,h({E,\bf r})
\label{Eq:CR_flux}
\end{equation}
where $\varphi_{\rm CR,\odot}(E)$ represents the local flux\footnote{In this work, we use the symbol $\varphi$ for angle-differential fluxes and $\Phi$ for angle-integrated fluxes.}, $g(r)$ is an adimensional function (normalized to one at the Sun position ${\bf r}_\odot$) introduced to describe the spatial distribution of CR while the function $h({E,\bf r})$, given by:
\begin{equation}
h(E,{\bf r})=\left(\frac{E}{\overline{E}}\right)^{\Delta({\bf r})}
\label{Eq:h_funct}
\end{equation}
with $\overline{E}=20\,{\rm GeV}$ and $\Delta({\bf r}_\odot)=0$, introduces a position-dependent variation $\Delta({\bf r})$ of the CR spectral index.

We describe the local CR nucleon flux $\varphi_{\rm CR,\odot}(E)$ according to the data driven parameterization \cite{Dembinski:2017} that relies as little as possible on theoretical assumptions. In the energy range of interest for the present analysis, that roughly corresponds to $E\simeq 10\times E_\gamma \sim 10-200\,{\rm TeV}$, the total nucleon flux is lower than the broken-power law parameterization \cite{Ahlers:2015} adopted in our previous calculation \cite{Pagliaroli:2016} by $\sim 12\,\%$ ($\sim 30 \,\%$) at $E\simeq 10\,$TeV ($E\simeq 100\,$TeV).

The function $g({\bf r})$ is determined by the distribution of the CR sources $f_{\rm S}({\bf r})$ in the Galaxy and by the propagation of CR in the Galactic magnetic field. In \cite{Pagliaroli:2016}, we considered two extreme assumptions, named {\em Case A} and {\em Case B}, that correspond to a completely mixed scenario (i.e. a uniform CR gas $g({\bf r})\equiv 1$) and to the opposite assumption of CR being confined very close to their sources (i.e. the function function $g({\bf r})$ was assumed to be proportional to the CR source density as reported in Eq.(3.9) of \cite{Pagliaroli:2016}), respectively.

Here, we consider a different phenomenological description. We define the function $g({\bf r})$ as:
\begin{equation}
     g({\bf r}) = \frac{1}{\mathcal{N}}\;\int d^3 x\; 
     f_{\rm S}({\bf r}-{\bf x})\;
     \frac{{\mathcal F}(|{\bf x}|/R)}{|{\bf x}|}
     \label{Eq:g_funct}
\end{equation} 
where $\mathcal{N}$ 
is a normalization constant given by:
\begin{equation}
     \mathcal{N} = \int d^3 x\; 
     f_{\rm S}({\bf r}_{\odot}-{\bf x})\;
     \frac{{\mathcal F}(|{\bf x}|/R)}{|{\bf x}|}
\end{equation} 
while the function ${\mathcal F}(\nu)$ is defined as:
\begin{equation}
  {\mathcal F}(\nu) \equiv \int_{\nu}^{\infty} d\gamma\; \frac{1}{\sqrt{2\pi}}\,\exp{\left(-{\gamma}^2/2\right)}
\end{equation} 
This kind of behaviour can be motivated as the solution of 3D isotropic diffusion equation with constant diffusion coefficient and stationary CR injection. In this context, the smearing parameter $R$ represents the diffusion length $R=\sqrt{2\,D\,t_{\rm G}}$, where $D$ is the diffusion coefficient and $t_{\rm G}$ is the integration time. The sources density $f_{\rm S}({\bf r})$ is assumed to follow the Supernova Remnants (SNR) number density parameterization given by \cite{Green:2015}.

The functions $g({\bf r})$ calculated for different smearing radii $R$  
are shown as a function of the galactocentric distance by the black lines in the left panel of Fig.\ref{fig:CR_dist} where they are compared with the CR density at $E\simeq 20\,{\rm GeV}$ and the $\gamma$-ray emissivity integrated above $E_\gamma=1 \,{\rm GeV}$ (which is a proxy of the CR flux) obtained from Fermi-LAT data in \cite{Acero:2016,Yang:2016,Pothast:2018}. 
%
While we believe that the data show a clear trend as a function of $r$, we think that they still do not allow to discard any of the proposed curves. We thus consider the two extreme assumptions $R=1\,{\rm kpc}$ and $R=\infty$, corresponding to the thick black lines in Fig.\ref{fig:CR_dist}, in order to encompass a large range of possibilities and to provide a conservative estimate of the uncertainty in the gamma-ray flux connected with different descriptions of the CR spatial distribution in the Galaxy\footnote{Since the assumed smearing length is larger than the thickness of the galactic disk, we neglect in both cases the variation of the CR flux along the galactic latitudinal axis. In other words, we assume $g(r,z)\simeq g(r,0)$ in the disk, where we used galactic cylindrical coordinates.}.

In the first case ($R=1\,{\rm kpc}$) the obtained function $g({\bf r})$ is very close to the SNR distribution \cite{Green:2015} given by the blue dashed line in Fig.\ref{fig:CR_dist}. However, differently from the SNR distribution, it does not vanish at the galactic center, as it is naturally expected due to CR propagation. In the outer regions of the Galaxy, the function $g({\bf r})$ drops faster than the CR density. For this reason, we neglect the variations of $g(r)$ and we assume that it is constant for $r\ge 10\,{\rm kpc}$.

In the second case ($R=\infty$) the function $g({\bf r})$ is quite close to the CR distribution predicted by the GALPROP code \cite{Galprop}, shown with orange dashed line in Fig.\ref{fig:CR_dist}. The function $g({\bf r})$ describes quite well the observational results a part from few points at $r\simeq 3\,{\rm kpc}$ and very close to the Galactic center where, however, results are less reliable for the reasons discussed in \cite{Pothast:2018}. 
 
\begin{figure}[!t]
\begin{center}
\includegraphics[width=8cm]{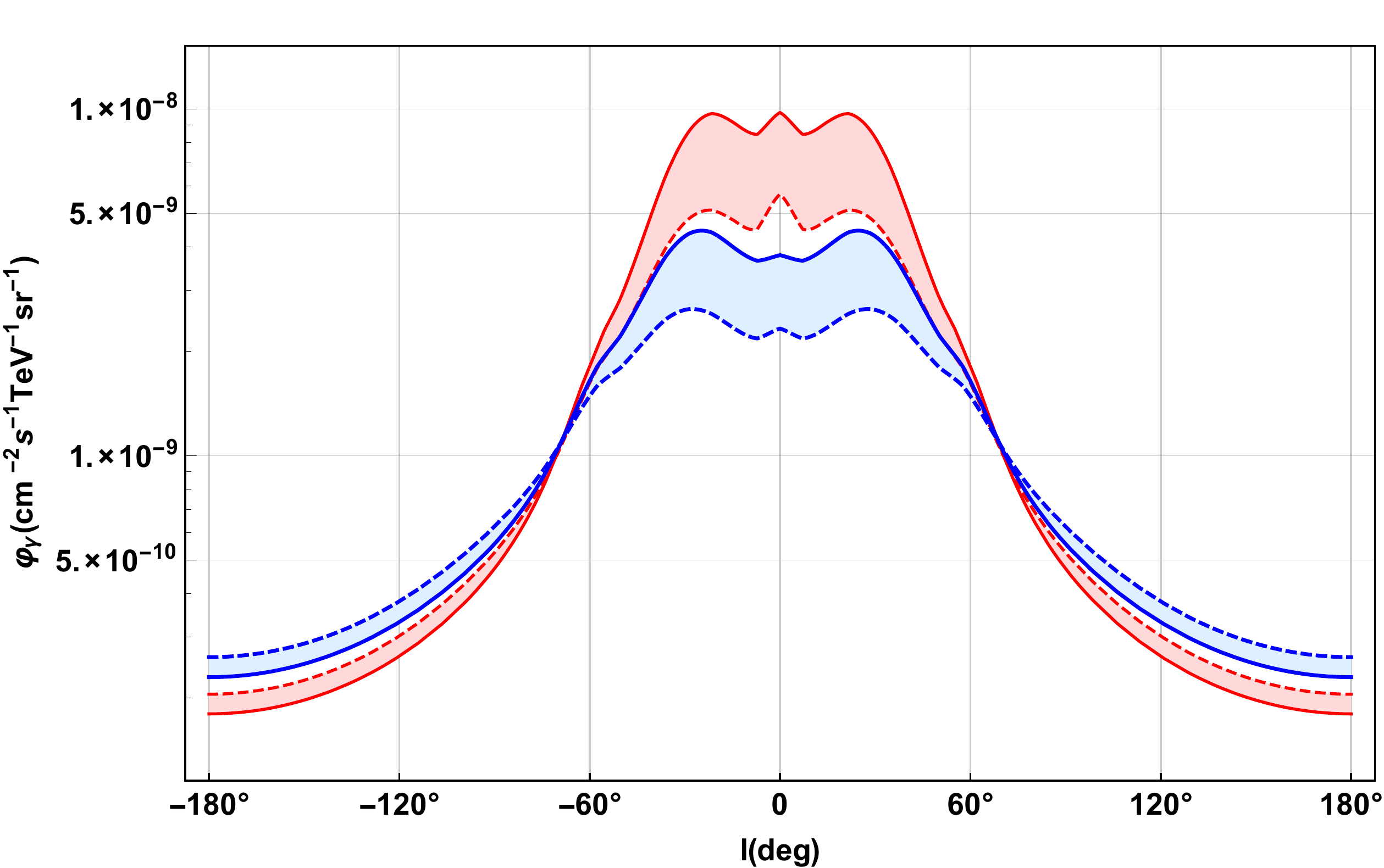}
\caption{\small\em 
The diffuse HE gamma-ray flux at $E_\gamma=1\,{\rm TeV}$ as a function of the Galactic longitude $l$ (for $b=0$) obtained by assuming that the CR spectrum is position-independent (blue lines) and by implementing CR spectral hardening in the inner Galaxy (red lines) as described by Eqs.(\ref{Eq:h_funct},\ref{Eq:Delta}). Solid and dashed lines (in each group) are obtained by assuming that the CR spatial distribution follows that of SNR with smearing radius $R=1\,{\rm kpc}$ and $R=\infty$, respectively. 
\label{fig:Gamma_dif}}
\end{center}
\end{figure}

The possibility of CR spectral hardening in the inner Galaxy is implemented by modelling the function $\Delta({\bf r})$ in Eq.(\ref{Eq:h_funct}) as:
\begin{equation}
\Delta(r,z)= \Delta_0\left(1 - \frac{r}{r_{\odot}}\right)
\label{Eq:Delta}
\end{equation} 
with $r_{\odot}=8.5\,$kpc, in galactic cylindrical coordinates. 
This choice is equivalent to what is done by \cite{Gaggero:2014,Gaggero:2015,Gaggero:2017jts} in their phenomenological CR propagation model characterized by radially dependent transport properties.
The numerical parameter $\Delta_0$ that physically corresponds to the difference between the CR spectral index at the Sun position, $\alpha_{\odot}\simeq 2.7$ at $E=20\,$GeV, and its value close to the galactic center, is taken as $\Delta_0=0.3$ since this assumption allows us to reproduce the trend with $r$ observed by \cite{Acero:2016,Yang:2016,Pothast:2018} for $r\le 10\,$kpc, as it is shown in the right panel of Fig.\ref{fig:CR_dist}\footnote{Refs. \cite{Yang:2016,Pothast:2018} report the spectral index $\alpha_\gamma$ of gamma emission associated to $\pi_0$ decay. This is converted into the spectral index $\alpha$ of the parent CR by $\alpha_\gamma \simeq \alpha + 0.1$, as it is discussed in \cite{Kappes:2006}.}.
In more external regions, the evidence for a variation of CR energy distribution is much weaker and we assume that $\Delta(r)$ is constant, as shown by the black solid line.

Our results for the diffuse HE gamma-ray flux produced by the interaction of CR with the gas contained in the Galactic disk at $E_\gamma = 1\,$TeV are plotted as function of the Galactic longitude $l$ (at the fixed latitude $b=0$) in Fig.\ref{fig:Gamma_dif}. Blue lines are obtained by assuming that the CR spectrum is independent from the position in the Galaxy (we refer to this in the following as the "standard" scenario) while red lines implement CR spectral hardening in the inner Galaxy, as described above. Solid and dashed lines in (each group) are obtained by assuming that the CR spatial distribution is described by the function $g(r)$ given in Eq.(\ref{Eq:g_funct}) with smearing radius equal to $R=1\,$kpc and $R=\infty$, respectively. 

The angle-integrated gamma-ray flux at $1\,$Tev is equal to $\Phi_\gamma = (7.0 - 8.0)\times 10^{-13}\;{\rm cm}^{-2}\;{\rm s}^{-1}\;{\rm GeV}^{-1}$  in the standard scenario with upper and lower bounds corresponding to $R=1\,$kpc and $R=\infty$, respectively. The inclusion of CR hardening increases the integrated flux by $1.2-1.3$, the exact enhancement factor being dependent on the assumed CR spatial distribution. Even if the effect on the total flux is relatively small, CR hardening may be responsible for a much more significant increase of the gamma-ray flux in the central region $-60^\circ \lapprox l \lapprox 60^\circ$. We see indeed from Fig.\ref{fig:Gamma_dif} that the enhancement factor can be as large as $\sim 2$ in the direction of the Galactic center. This factor is larger than the uncertainty in the flux due to CR spatial distribution that we conservatively represent as the width of the coloured bands in the figure. 
Additional uncertainty sources are the normalization of the local CR flux at $E\sim 10\, E_\gamma \sim 10\,$TeV, the distribution of gas in the Galaxy for which we use the same input as \cite{Galprop}\footnote{The heavy element contribution is included by assuming that the total mass of the interstellar gas is a factor $1.42$ larger that the mass of hydrogen, as it is expected if the solar system composition is considered representative for the entire galactic disk.}, the hadronic interaction cross section, etc.
All these are expected to produce a total error smaller than the difference between red and blue lines in the figure,  
suggesting that effects of CR hardening can be probed by TeV scale gamma-ray observations of the galactic central region.

\begin{figure}[t!]
\begin{center}
\includegraphics[width=8cm]{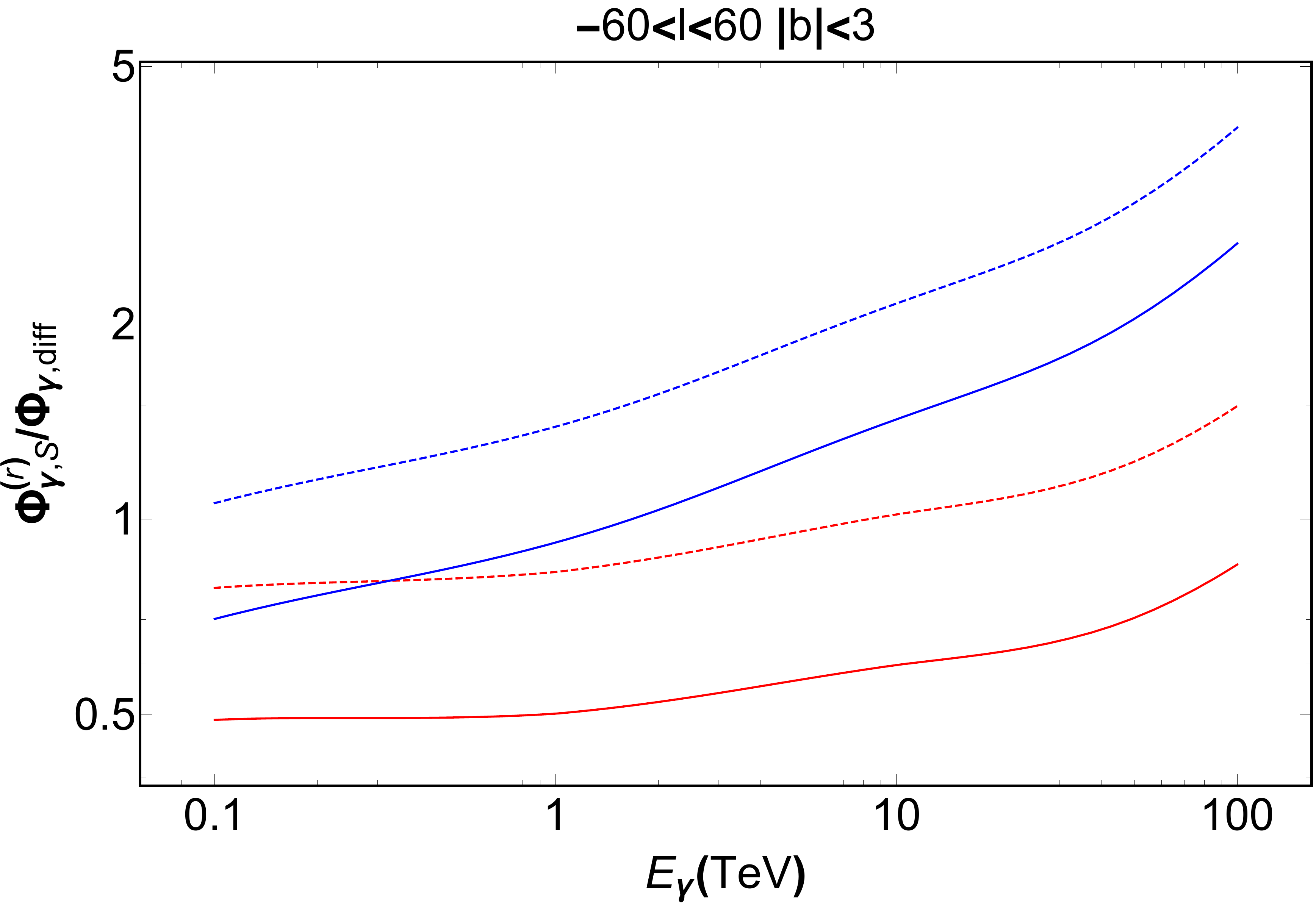}
\caption{\small\em The ratio between the cumulative flux produced by sources in the HESS-GPS catalogue and the diffuse emission in the Galactic region $-60^\circ \le l \le 60^\circ$ and $-3^\circ \le b \le 3^\circ$ as a function of the gamma energy. Blue lines correspond to the "standard" scenario while red lines implement CR spectral hardening. Solid (dashed) lines are obtained by assuming a smearing radius equal to $R=1\,$kpc ($R=\infty$). \label{fig:ratio}}
\end{center}
\end{figure}

\section{The sources contribution}
\label{sec2}

The total flux of HE gammas produced in our Galaxy can be written as:
\begin{equation}
\varphi_{\gamma,\rm tot} = \varphi_{\gamma, \rm diff} + \varphi_{\gamma,\rm S} + \varphi_{\gamma,\rm IC}
\end{equation}
where $\varphi_{\gamma, \rm diff}$ is the diffuse flux, $\varphi_{\gamma,\rm S}$ is the flux produced by sources and $\varphi_{\gamma,\rm IC}$ indicates the contribution due to inverse Compton emission by diffuse HE electrons. 
With the term 'sources', we refer here to all the contributions produced within or close to an acceleration site by freshly accelerated particles that potentially have (a part from cut-off effects) harder spectra than the diffuse component. 
In the longitude range of interest ($-60^\circ \lapprox l \lapprox 60^\circ$), the gamma-ray sky gets a comparable contribution at TeV energies by sources and diffuse emission and it is difficult to isolate the two components. For this reason, we construct a complete model that include both contributions 
and we compare it with the total flux observed by the various gamma-ray experiments. 

The diffuse emission is calculated as described in the previous section. 
The flux produced by resolved sources, which will be indicated in the following as $\varphi^{(r)}_{\gamma,S}$, is evaluated by using the HESS-GPS catalogue \cite{HESS:2018} that includes 78 VHE sources observed in the longitude range $-110^\circ \le l \le 60^\circ$ and latitudes $|b|<3^\circ$, measured with an angular resolution of $0.08^\circ$ and a sensitivity $\lsim 1.5\%$ Crab flux for point-like objects. 
%
Extended sources are described with Gaussian profiles in $l$ and in $b$, centered at the source coordinates and with the size $\sigma$ reported in the catalogue. 
The source spectrum is calculated in the energy range $0.1 \lapprox E_\gamma \lapprox 100\,$TeV by using a power law or a power-law with an exponential cutoff with the best-fit parameters reported in the online material of \cite{HESS:2018}.
The cumulative flux produced by observed sources in the angular region $-60^\circ < l < 60^\circ$ and $-3^\circ< b < 3^\circ$ is compared with diffuse emission in Fig.\ref{fig:ratio}. We see that it corresponds to a fraction $50-150\%$ of the diffuse flux at $E_\gamma\simeq 1\,$TeV, depending on the assumed scenario for CR space and energy distribution. The source contribution becomes even more relevant at larger energy and dominates the total emission at $E_\gamma \sim 100$ TeV.

We remark that the flux $\varphi^{(r)}_{\rm S}$, includes by construction only resolved sources. The total flux produced by sources should include also the contribution due to very faint and/or extended sources that are not resolved by HESS. We can express the total source contribution as
\begin{equation}
\varphi_{\gamma,\rm S} = \varphi^{(r)}_{\gamma,\rm S} \; (1 + \eta)
\label{eta}
\end{equation}
where the parameter $\eta$, that depends in principle on the observation direction and energy, quantifies the relative contribution of unresolved objects to the total source flux. 
\begin{figure*}[!t]
\begin{center}
\includegraphics[width=8cm]{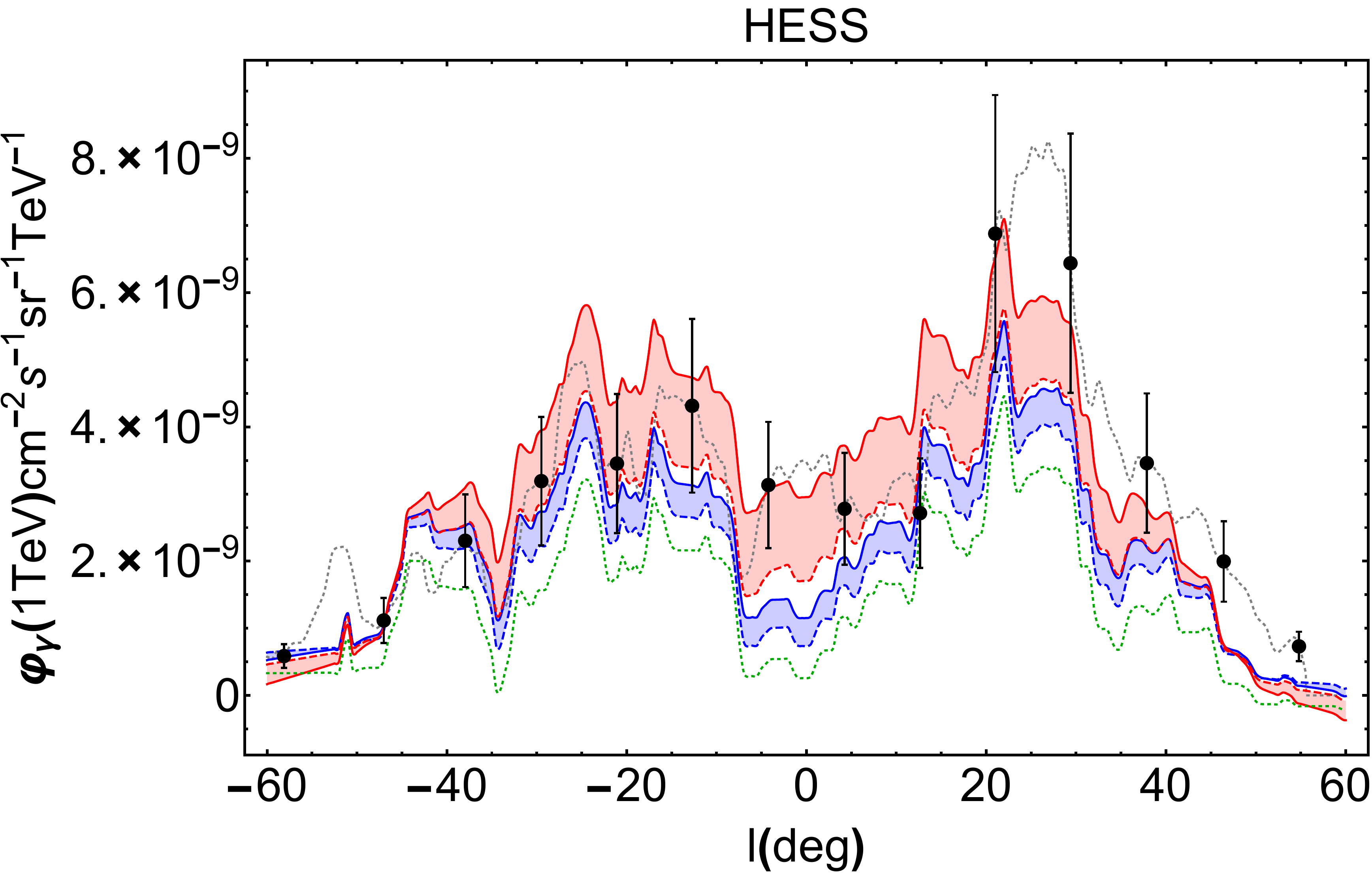}
\includegraphics[width=8cm]{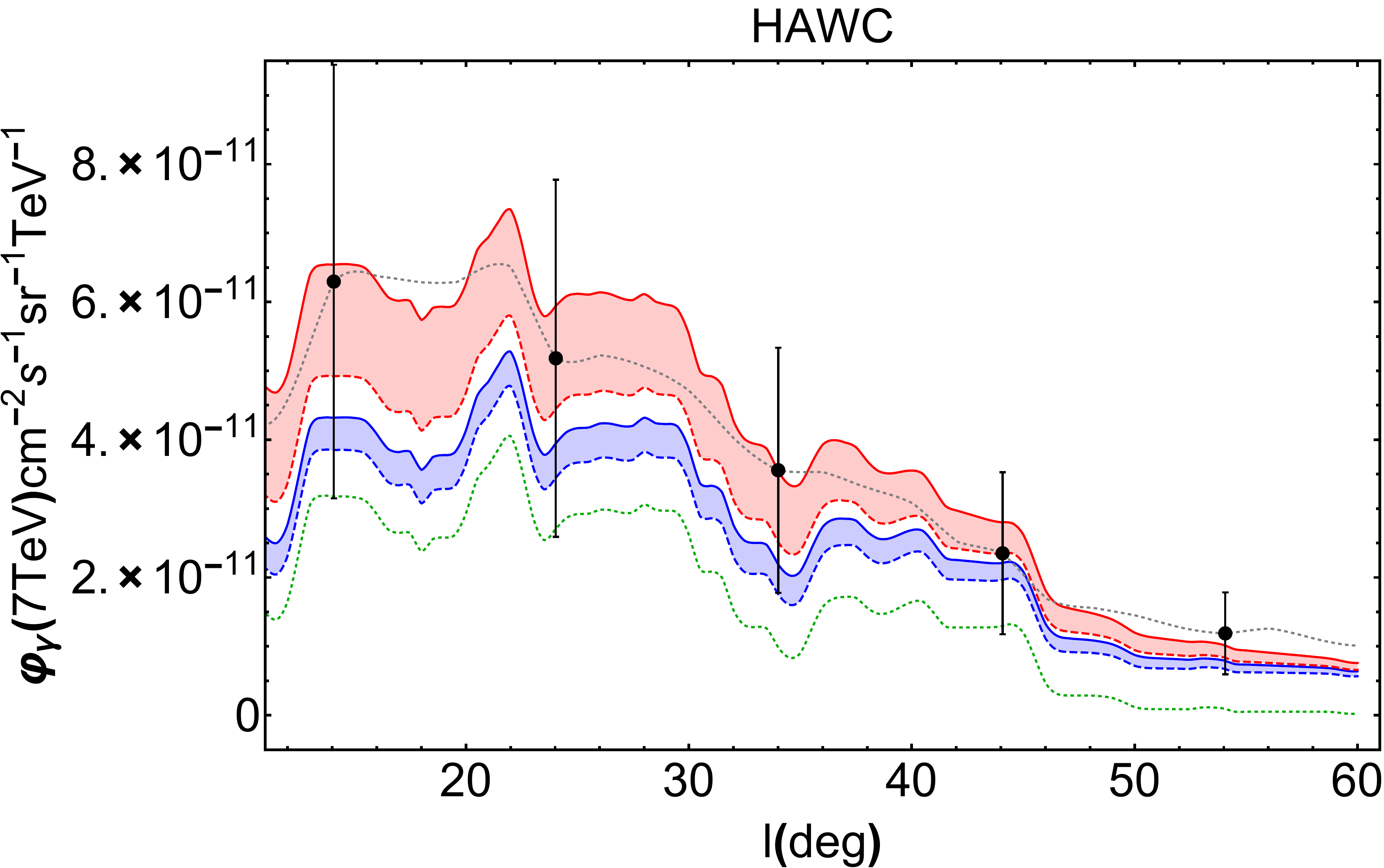}\\
\includegraphics[width=8cm]{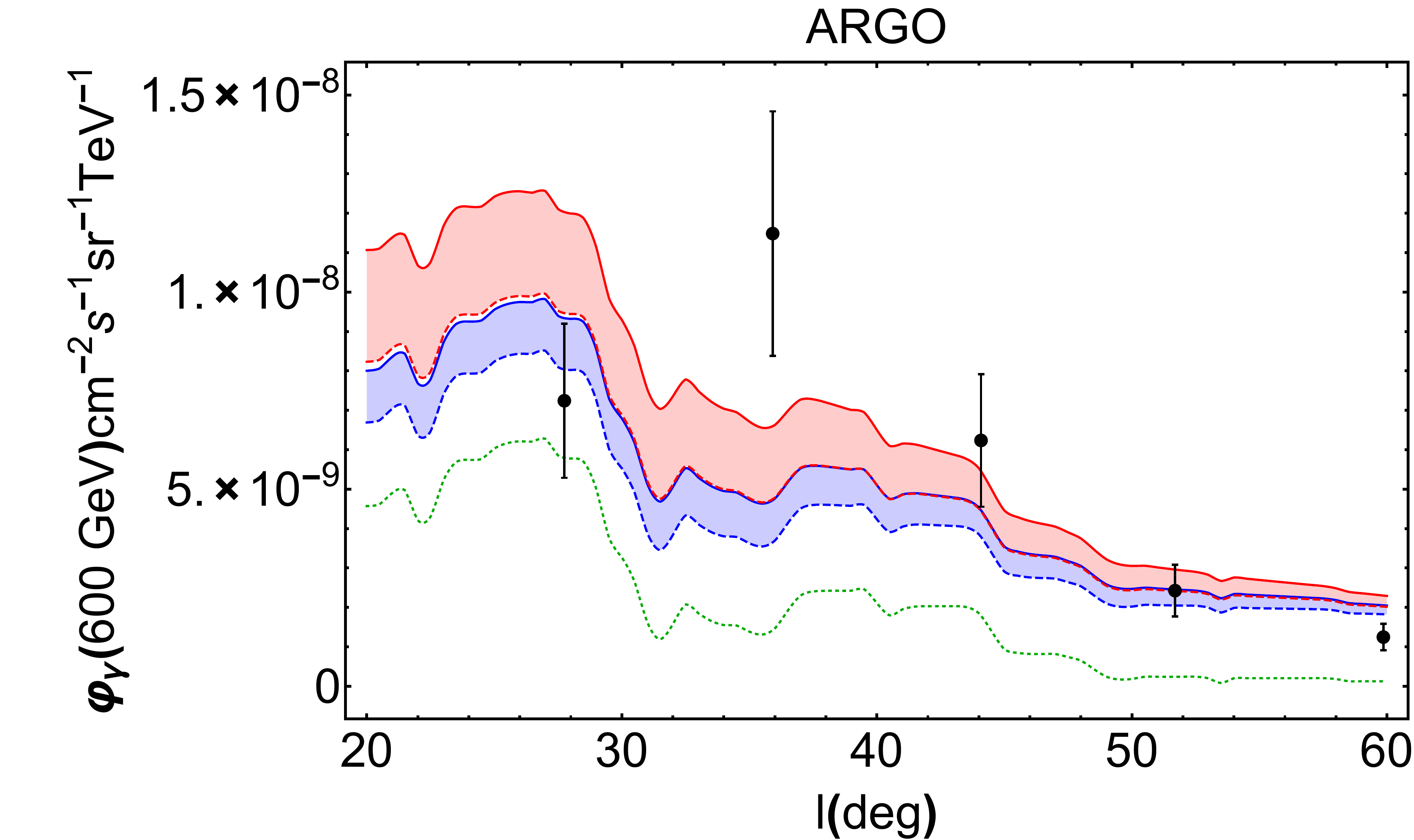}
\includegraphics[width=8cm]{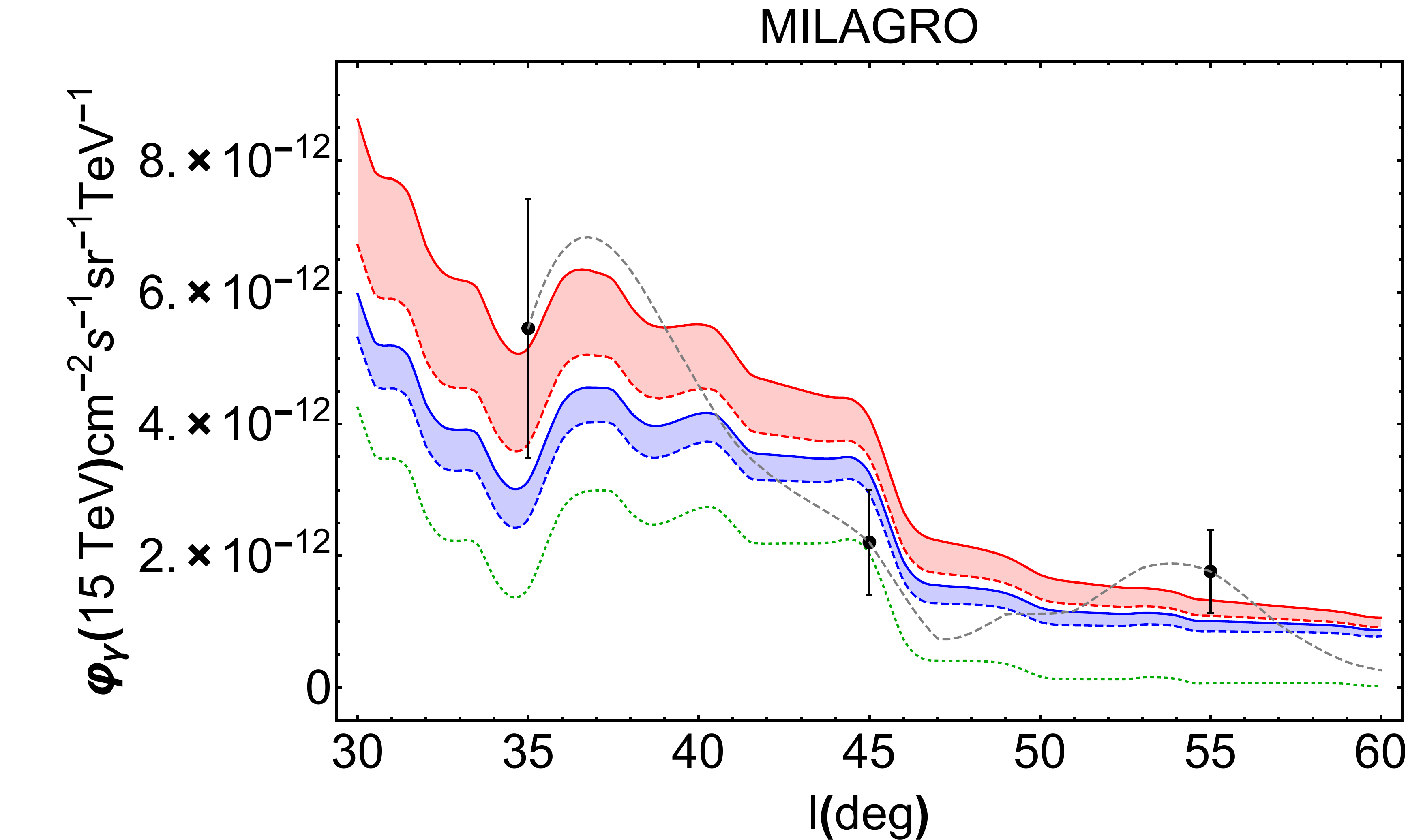}
\caption{\small\em The total gamma fluxes predicted as a function of the Galactic longitude $l$ following the same color prescription as in Fig.\ref{fig:Gamma_dif}. The green dotted line represents the contribution of resolved sources. For a detailed description of the plots see the text.
\label{fig:comp_with_data}}
\end{center}
\end{figure*} 
%
%
%
At TeV energies, we naturally expect that unresolved sources accounts for a relevant fraction of the total signal because a small portion of the Galaxy is resolved by the HESS experiment while sources are expected to be distributed in the entire Galaxy. 
It is e.g. discussed in \cite{HESS:2018} that point-like objects with intrinsic luminosity above 1 TeV equal to $L=10^{34}\,{\rm erg}\,{\rm s}^{-1}$ can be observed up to a median distance $\overline{D}\sim 7 \,{\rm kpc}$. The observational horizon can be, however, much smaller (up to a factor 3) for extended sources.  

By taking into account that HESS sensitivity limit for extended objects is $\simeq 10\%$ of the CRAB flux integrated above 1 TeV \cite{HESS:2018}, we can provide an estimate of the $\eta$ parameter as a function of the assumed sources intrinsic luminosity $L$, as it is described the Appendix \ref{App:UnresolvedSources}.
The integrated contribution of unresolved sources in the angular region $|l|\le 60^\circ$ and $|b|\le 3^\circ$ is found to be $\eta \simeq 60\%$ of the resolved signal, if we assume that source distribution is proportional to the SNR number density \cite{Green:2015} and sources have intrinsic luminosity $L=L_{CRAB}$.
This prediction is affected by large uncertainties and cannot be obtained as a function of the observation direction and energy. For this reason, we do not include the unresolved sources contribution when comparing of observational data of Argo-YBJ, HESS, HAWC and Milagro experiments with our theoretical estimates.   
However, the implications of unresolved sources for the interpretation of our results and the constraints on the parameter $\eta$ arising from observational data are considered in Sect.\ref{Sec:Discussion}.

\section{Comparison with data}
\label{sec3}

The sum of the flux produced by diffuse component and resolved sources: 
\begin{equation}
\varphi_{\gamma} =  \varphi_{\gamma, \rm diff} +\varphi^{(r)}_{\gamma, \rm S}
\label{Eq:phi_tot}
\end{equation}
provides a lower estimate of the total emission that can be compared with the observational determinations of the total gamma-ray flux at TeV energies obtained by HESS \cite{Abramowski:2014}, HAWC \cite{HAWCdata},  Argo-YBJ \cite{Bartoli:2015} and Milagro \cite{Abdo:2008}.
This comparison is reported in Fig.\ref{fig:comp_with_data}.

The four experiments have different observation windows, angular resolutions and median observation energies, as it is indicated in the figure. 
In particular, the HESS Galactic Plane Survey \cite{Abramowski:2014} (upper left panel of Fig.\ref{fig:comp_with_data}) measures the profile of gamma-ray emission, averaged for latitudes $|b| < 2^\circ$, in the longitude range  $-75^\circ < l < 60^\circ$, for a photon median energy $E_\gamma = 1\,$TeV. 
The data are provided in angular bins $\delta l \sim 0.6^\circ$. 
In order to remove possible background contributions, the total emission from the Galactic plane is obtained by HESS as the excess with respect to the average signal at absolute latitudes $|b|\ge 1.2^\circ$ \cite{Abramowski:2014}.
We apply the same background reduction procedure to our predictions. For each considered case, we report in the figure the excess along the galactic plane (i.e. in the region $|b|<1.2^\circ$) with respect to the average emission in the region $1.2^\circ <|b|<2^\circ$.

The HAWC experiment recently presented the preliminary longitudinal gamma-ray profile in the angular region $0^\circ < l < 60^\circ$ for a photon median energy $E_\gamma = 7$ TeV (upper right panel of Fig.\ref{fig:comp_with_data}). The emission is averaged over galactic latitudes $|b| < 2^\circ$ and it is provided in angular bins $\delta l \sim 2^\circ$ \cite{HAWCdata}.
The Argo-YBJ experiment measures the total gamma-ray emission 
in the overlapping longitudinal range $ 25 ^\circ < l < 60^\circ$
for $E_\gamma=600$ GeV (lower left panel of Fig.\ref{fig:comp_with_data}). The measured flux is averaged over galactic latitudes $|b| < 5^\circ$ and the data are presented in longitudinal bins $\delta l \sim 8^\circ$ \cite{Bartoli:2015}.  
At higher energy, $E_\gamma = 15$ TeV, the Milagro experiment reports the total gamma-ray emission in the region $30^\circ < l < 60^\circ$ (lower right panel of Fig.\ref{fig:comp_with_data}). The observed flux is averaged over galactic latitudes $|b| < 2^\circ$ and it is provided in angular bins $\delta l \sim 2^\circ$\cite{Abdo:2008}.  

In order to facilitate the comparison among the different measurements, we re-binned the data from HESS, HAWC and Milagro over longitudinal bins $\delta l \sim 10^\circ$ that are comparable with Argo-YBJ angular resolution.
The re-binning procedure is also done to avoid large fluctuations of the signal thus making visually clear the excess (or the deficit) of the observed flux with respect to our predictions.
It should be noted that our calculations of the diffuse gamma-ray component are intended to describe the main features of the interstellar emission, but they are not expected to reproduce the small angular scales fluctuations of the observed signal, being based on (relatively) smooth gas and CR distributions.
The data points obtained after the re-binning procedure are reported in Fig.\ref{fig:comp_with_data}. They are located in the bin midpoint and have vertical errors bars that include systematic and statistical errors of the measured fluxes. 
The two errors are summed in quadrature and correspond to a total uncertainty equal to $30\%$ for HESS \cite{HESS:2018}, $50\%$ for HAWC \cite{HAWCdata}, $27\%$ for Argo \cite{Bartoli:2015} and $36\%$ for Milagro \cite{Abdo:2008}. 
The grey dashed lines (not given for Argo-YBJ) are obtained by performing a moving average of the experimental data with the previously discussed $\delta l$ and are intended to show the longitudinal dependence of the total galactic emission measured by each experiment.

The red and blue curves displayed in Fig.\ref{fig:comp_with_data} give the sum of the gamma-ray flux produced by sources in HESS catalogue and diffuse emission, averaged in the galactic latitudinal windows considered by each experiment and then smeared in longitudinal bins $\delta l= 10^\circ$ for HESS, HAWC and Milagro and $\delta l=8^\circ$ for Argo-YBJ.
The red lines take into account CR spectral hardening hypothesis 
while blue lines correspond to the standard assumption. 
Solid and dashed lines are obtained (in both cases) by assuming that the CR spatial distribution follows that of SNRs with smearing radius equal to $R=1\,$kpc and $R=\infty$ (see Eq.\ref{Eq:g_funct}), respectively. The green dotted lines in the various panels show the contribution to the total flux which is only ascribed to sources resolved by HESS. 

Finally, in order to discuss the behaviour of our predictions as a function of energy, we select the observation window $30^\circ < l < 60^\circ$ and $-2^\circ< b < 2^\circ$ where all the experiments are sensitive. In this region, we compare the gamma-ray fluxes obtained in this work with the measured ones at different energies. The results are reported in Fig.\ref{fig:spectra}.  

\section{Results}
\label{Sec:Discussion}

\begin{figure}[!t]
\begin{center}
\includegraphics[width=8cm]{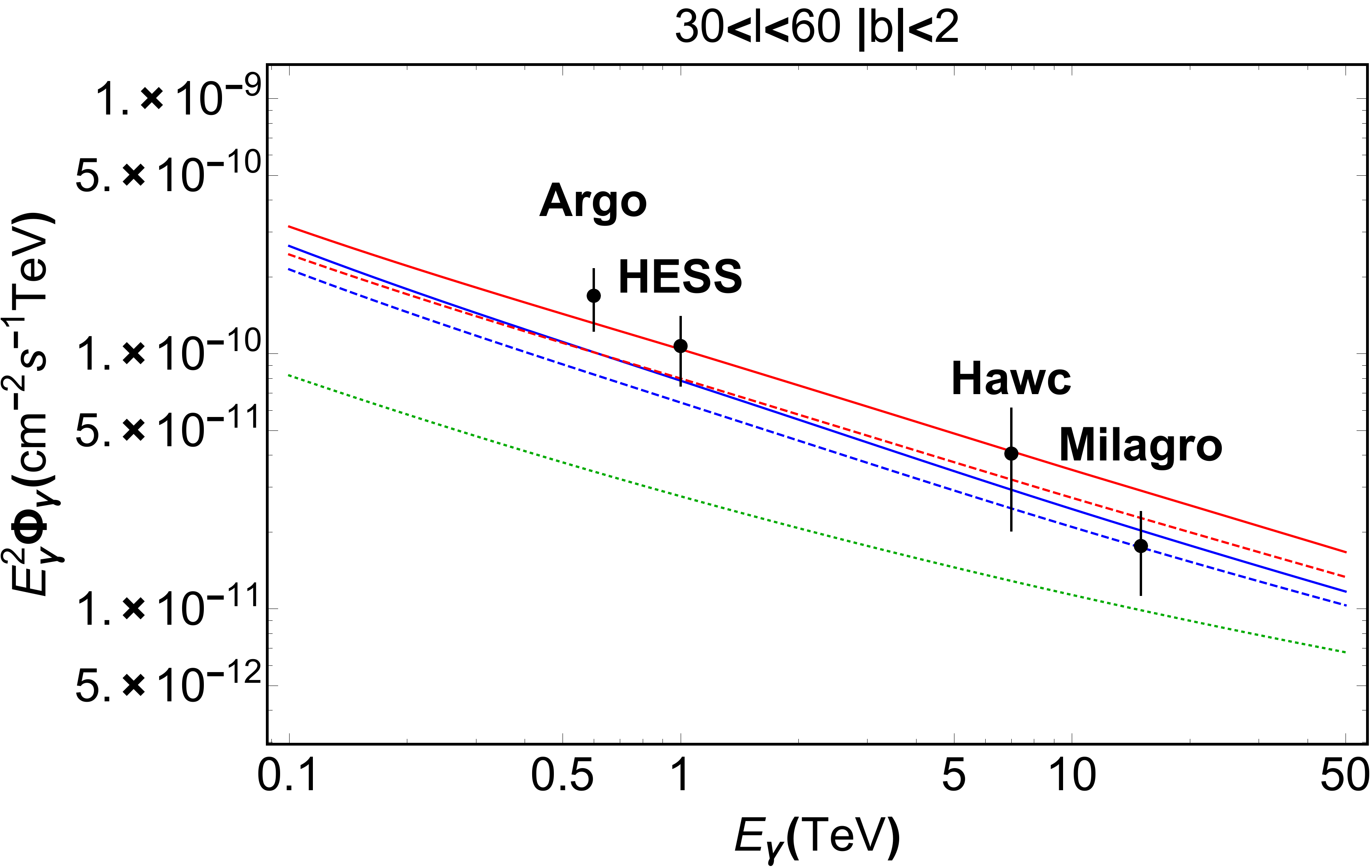}
\caption{\small\em The total gamma flux as a function of the energy for the different cases considered in this work. The same quantity as measured by HESS, HAWC, Argo-YBJ and Milagro is also displayed for a comparison. The green dotted line represents the contribution due to resolved sources.
\label{fig:spectra}}
\end{center}
\end{figure}

The results reported in Fig.\ref{fig:comp_with_data} and \ref{fig:spectra} allow us to reach several important conclusions.

{\em i)} The main features of the gamma-ray signal in the TeV domain are reproduced by our model both as function of the observation angle and energy. 
The possibility to validate our approach and to benchmark our calculations with gamma-ray data is particularly important to predict at larger energies $E_\nu\sim 100\,$TeV, within a multi-messenger approach, the galactic contribution to neutrino telescopes signal, see e.g.\cite{Pagliaroli:2017}.

{\em ii)} According to our estimates, the emission from resolved sources accounts for a large fraction of the total flux in the TeV domain. 
In the angular ranges indicated in Fig.\ref{fig:comp_with_data}, the gamma-ray sources included in the HESS-GPS catalogue are responsible for $30\%$, $46\%$, $44\%$, $44\%$ of the total signal observed by Argo-YBJ, HESS, HAWC and Milagro, respectively.
Note that gamma-ray sources, if powered by hadronic processes, are also emitting HE neutrinos. Our calculations show that their contribution 
to galactic neutrino fluxes is potentially large and thus cannot be neglected, as was also suggested in \cite{Pagliaroli:2017}.   

{\em iii)} It was noted in \cite{Pagliaroli:2017} that gamma-ray flux from the galactic plane at $E_\gamma =1\,$TeV is not symmetric with respect to galactic center, being the flux observed by HESS in the longitude range $0^\circ\le l \le 60^\circ$ about $30\%$ larger than that observed in $-60^\circ \le l \le 0^\circ$.
This asymmetry is partially reproduced by our calculations as a result of the inclusion of resolved sources. In our model, diffuse emission is indeed invariant with respect to $l\to -l$ transformation, being based on cylindrically symmetric gas and CR distribution. 
In order to reproduce the observed difference, one should break cylindrical symmetry, including e.g. a 3D gas distribution and/or clumps of increasing CR density at fixed galactic coordinates \cite{Aharonian:2018rob}.

{\em iv)} The comparison between theoretical predictions and observational data allows us to obtain interesting constraints on the CR distribution.
Indeed, the predicted fluxes in the presence of CR hardening saturate or exceed the observed signal in certain regions of the Sky. 
In particular, the largest prediction considered in this work, which is obtained by including CR spectral hardening and by assuming that the CR spatial distribution follows that of SNR with $R=1\,$kpc smearing radius (red solid lines), account for $95\%$, $104\%$, $104\%$ and $118\%$ of the total signal observed by Argo-YBJ, HESS, HAWC and Milagro in the angular ranges indicated in Fig.\ref{fig:comp_with_data}, respectively.
This is hardly acceptable on physical basis considering that an additional non negligible contribution is  expected to be provided by non-resolved sources and inverse Compton.

{\em v)} The contribution from inverse Compton and unresolved sources can be estimated by subtracting theoretical predictions from observational data, allowing us to obtain an upper limit on the $\eta$ parameter that quantifies the relative contribution of unresolved objects to the total source flux, see Eq.(\ref{eta}).
If we consider the lowest prediction in the presence of CR spectral hardening, that is obtained by considering a smooth CR space distribution (i.e. with a smearing radius $R=\infty$) and corresponds to the red dashed lines in Fig.\ref{fig:comp_with_data}, we obtain that $\eta$ should be lower than $0.92$, $0.51$, $0.45$ and $0.14$ in the angular ranges and at the gamma-ray energies probed by Argo-YBJ, HESS, HAWC and Milagro, respectively. 
Note that the IC contribution to HESS signal is expected to be strongly reduced \cite{Abramowski:2014} due to the very stringent latitudinal cut for background suppression procedure. Thus the HESS data can be used to extract the unresolved source contribution at $E_\gamma=1\,{\rm TeV}$, obtaining $\eta \le 0.5$ in the presence of CR hardening.

{\em vi)} The obtained limits for unresolved sources contribution can be compared with the expected values for $\eta$ with potential implications for source population studies. It is e.g. estimated in Appendix \ref{App:UnresolvedSources} that $\eta \sim 0.6$ for photon energies $E_\gamma\ge 1\,{\rm TeV}$ and in the angular range $|l|\le 60^\circ$ and $|b|\le 3^\circ$ 
that essentially corresponds to the region probed by HESS total flux measurement, see upper-left panel of Fig.\ref{fig:comp_with_data}.
This estimate is obtained by assuming that HE gamma-ray sources are distributed in the Galaxy as SNRs \cite{Green:2015} and have an intrinsic luminosity comparable to that of CRAB nebula. 
Larger values for $\eta$ are obtained for smaller sources intrinsic luminosity, since the horizon for resolved sources decreases (see Eq.\ref{Eq:horizon}) and thus a smaller portion of the Galaxy is probed by HESS.     

{\em vii)} By comparing the estimated value for $\eta$ with the constraint $\eta \le 0.5$ obtained from HESS results, we are lead to the conclusion that CR spectral hardening is consistent with observational data if we assume that HE gamma-ray sources have a relatively large intrinsic luminosity, greater than or comparable to that of CRAB.
On the contrary, theoretical arguments and/or observational indications for a large unresolved sources contribution (i.e. $\eta \gg 0.5$) could rule out the hypothesis of CR spectral hardening, since the predicted signal in this assumption from diffuse and (resolved plus unresolved) sources would exceed the total observed gamma-ray flux.

\section{Summary}
\label{summary}

In this work, we demonstrate that the TeV gamma-ray sky can be used to probe the distribution of galactic cosmic rays. 
We present updated calculations of the diffuse gamma-ray emission produced by the interaction of CR with the gas contained in the galactic disk considering also the possibility of CR spectral hardening in the inner Galaxy, recently emerged from the analysis of Fermi-LAT data at lower energies. 
We show that interesting constraints on the CR distribution in the Galaxy can be obtained by comparing the total flux due to the diffuse interstellar emission plus the cumulative contribution of point-like and extended sources in HESS-GPS catalogue with the total flux measured by Argo-YBJ, HESS, HAWC and Milagro experiments. 
Indeed, our largest prediction for the diffuse emission which is obtained 
by assuming that the CR density follows that of SNR (with $R=1\,$kpc smearing radius) and by considering CR spectral hardening in the inner Galaxy, seems to be excluded by present observational data.
Stronger constraints can be obtained by taking into account the contribution to the observed signal provided by sources non resolved by HESS. This is naturally expected to be non negligible but is affected by large uncertainties.
The CR spectral hardening hypothesis would be ruled out globally (i.e. even if a smooth CR spatial distribution with $R=\infty$ is assumed), if the non resolved contribution is demonstrated to be larger than $\sim 50\%$ of the resolved component.

\acknowledgments
This work was partially supported by the research grant number 2017W4HA7S ``NAT-NET: Neutrino and Astroparticle Theory Network'' under the program PRIN 2017 funded by the Italian Ministero dellÕIstruzione, dell'Universit\`a e della Ricerca (MIUR). 
   
\bibliography{HE-nu-gamma-astronomy}

\begin{thebibliography}{29}
\expandafter\ifx\csname natexlab\endcsname\relax\def\natexlab#1{#1}\fi
\expandafter\ifx\csname bibnamefont\endcsname\relax
  \def\bibnamefont#1{#1}\fi
\expandafter\ifx\csname bibfnamefont\endcsname\relax
  \def\bibfnamefont#1{#1}\fi
\expandafter\ifx\csname citenamefont\endcsname\relax
  \def\citenamefont#1{#1}\fi
\expandafter\ifx\csname url\endcsname\relax
  \def\url#1{\texttt{#1}}\fi
\expandafter\ifx\csname urlprefix\endcsname\relax\def\urlprefix{URL }\fi
\providecommand{\bibinfo}[2]{#2}
\providecommand{\eprint}[2][]{\url{#2}}

\bibitem[{\citenamefont{{Kraushaar} et~al.}(1972)\citenamefont{{Kraushaar},
  {Clark}, {Garmire}, {Borken}, {Higbie}, {Leong}, and
  {Thorsos}}}]{Kraushaar:1972}
\bibinfo{author}{\bibfnamefont{W.~L.} \bibnamefont{{Kraushaar}}},
  \bibinfo{author}{\bibfnamefont{G.~W.} \bibnamefont{{Clark}}},
  \bibinfo{author}{\bibfnamefont{G.~P.} \bibnamefont{{Garmire}}},
  \bibinfo{author}{\bibfnamefont{R.}~\bibnamefont{{Borken}}},
  \bibinfo{author}{\bibfnamefont{P.}~\bibnamefont{{Higbie}}},
  \bibinfo{author}{\bibfnamefont{V.}~\bibnamefont{{Leong}}}, \bibnamefont{and}
  \bibinfo{author}{\bibfnamefont{T.}~\bibnamefont{{Thorsos}}},
  \bibinfo{journal}{\apj} \textbf{\bibinfo{volume}{177}}, \bibinfo{pages}{341}
  (\bibinfo{year}{1972}).

\bibitem[{\citenamefont{{Kniffen} et~al.}(1973)\citenamefont{{Kniffen},
  {Hartman}, {Thompson}, and {Fichtel}}}]{Kniffen:1973}
\bibinfo{author}{\bibfnamefont{D.~A.} \bibnamefont{{Kniffen}}},
  \bibinfo{author}{\bibfnamefont{R.~C.} \bibnamefont{{Hartman}}},
  \bibinfo{author}{\bibfnamefont{D.~J.} \bibnamefont{{Thompson}}},
  \bibnamefont{and} \bibinfo{author}{\bibfnamefont{C.~E.}
  \bibnamefont{{Fichtel}}}, \bibinfo{journal}{\apjl}
  \textbf{\bibinfo{volume}{186}}, \bibinfo{pages}{L105} (\bibinfo{year}{1973}).

\bibitem[{\citenamefont{{Mayer-Hasselwander}
  et~al.}(1982)\citenamefont{{Mayer-Hasselwander}, {Bennett}, {Bignami},
  {Buccheri}, {Caraveo}, {Hermsen}, {Kanbach}, {Lebrun}, {Lichti}, {Masnou}
  et~al.}}]{Mayer-Hasselwander:1982}
\bibinfo{author}{\bibfnamefont{H.~A.} \bibnamefont{{Mayer-Hasselwander}}},
  \bibinfo{author}{\bibfnamefont{K.}~\bibnamefont{{Bennett}}},
  \bibinfo{author}{\bibfnamefont{G.~F.} \bibnamefont{{Bignami}}},
  \bibinfo{author}{\bibfnamefont{R.}~\bibnamefont{{Buccheri}}},
  \bibinfo{author}{\bibfnamefont{P.~A.} \bibnamefont{{Caraveo}}},
  \bibinfo{author}{\bibfnamefont{W.}~\bibnamefont{{Hermsen}}},
  \bibinfo{author}{\bibfnamefont{G.}~\bibnamefont{{Kanbach}}},
  \bibinfo{author}{\bibfnamefont{F.}~\bibnamefont{{Lebrun}}},
  \bibinfo{author}{\bibfnamefont{G.~G.} \bibnamefont{{Lichti}}},
  \bibinfo{author}{\bibfnamefont{J.~L.} \bibnamefont{{Masnou}}},
  \bibnamefont{et~al.}, \bibinfo{journal}{\aap} \textbf{\bibinfo{volume}{105}},
  \bibinfo{pages}{164} (\bibinfo{year}{1982}).

\bibitem[{\citenamefont{{Hunter} et~al.}(1997)\citenamefont{{Hunter},
  {Bertsch}, {Catelli}, {Dame}, {Digel}, {Dingus}, {Esposito}, {Fichtel},
  {Hartman}, {Kanbach} et~al.}}]{Hunter:1997}
\bibinfo{author}{\bibfnamefont{S.~D.} \bibnamefont{{Hunter}}},
  \bibinfo{author}{\bibfnamefont{D.~L.} \bibnamefont{{Bertsch}}},
  \bibinfo{author}{\bibfnamefont{J.~R.} \bibnamefont{{Catelli}}},
  \bibinfo{author}{\bibfnamefont{T.~M.} \bibnamefont{{Dame}}},
  \bibinfo{author}{\bibfnamefont{S.~W.} \bibnamefont{{Digel}}},
  \bibinfo{author}{\bibfnamefont{B.~L.} \bibnamefont{{Dingus}}},
  \bibinfo{author}{\bibfnamefont{J.~A.} \bibnamefont{{Esposito}}},
  \bibinfo{author}{\bibfnamefont{C.~E.} \bibnamefont{{Fichtel}}},
  \bibinfo{author}{\bibfnamefont{R.~C.} \bibnamefont{{Hartman}}},
  \bibinfo{author}{\bibfnamefont{G.}~\bibnamefont{{Kanbach}}},
  \bibnamefont{et~al.}, \bibinfo{journal}{\apj} \textbf{\bibinfo{volume}{481}},
  \bibinfo{pages}{205} (\bibinfo{year}{1997}).

\bibitem[{\citenamefont{Ackermann et~al.}(2012)}]{Ackermann:2012}
\bibinfo{author}{\bibfnamefont{M.}~\bibnamefont{Ackermann}}
  \bibnamefont{et~al.} (\bibinfo{collaboration}{Fermi-LAT}),
  \bibinfo{journal}{Astrophys. J.} \textbf{\bibinfo{volume}{750}},
  \bibinfo{pages}{3} (\bibinfo{year}{2012}), \eprint{1202.4039}.

\bibitem[{\citenamefont{Acero et~al.}(2016)}]{Acero:2016}
\bibinfo{author}{\bibfnamefont{F.}~\bibnamefont{Acero}} \bibnamefont{et~al.}
  (\bibinfo{collaboration}{Fermi-LAT}), \bibinfo{journal}{Astrophys. J. Suppl.}
  \textbf{\bibinfo{volume}{223}}, \bibinfo{pages}{26} (\bibinfo{year}{2016}),
  \eprint{1602.07246}.

\bibitem[{\citenamefont{Gaggero
  et~al.}(2015{\natexlab{a}})\citenamefont{Gaggero, Grasso, Marinelli, Urbano,
  and Valli}}]{Gaggero:2015}
\bibinfo{author}{\bibfnamefont{D.}~\bibnamefont{Gaggero}},
  \bibinfo{author}{\bibfnamefont{D.}~\bibnamefont{Grasso}},
  \bibinfo{author}{\bibfnamefont{A.}~\bibnamefont{Marinelli}},
  \bibinfo{author}{\bibfnamefont{A.}~\bibnamefont{Urbano}}, \bibnamefont{and}
  \bibinfo{author}{\bibfnamefont{M.}~\bibnamefont{Valli}},
  \bibinfo{journal}{Astrophys. J.} \textbf{\bibinfo{volume}{815}},
  \bibinfo{pages}{L25} (\bibinfo{year}{2015}{\natexlab{a}}),
  \eprint{1504.00227}.

\bibitem[{\citenamefont{Yang et~al.}(2016)\citenamefont{Yang, Aharonian, and
  Evoli}}]{Yang:2016}
\bibinfo{author}{\bibfnamefont{R.}~\bibnamefont{Yang}},
  \bibinfo{author}{\bibfnamefont{F.}~\bibnamefont{Aharonian}},
  \bibnamefont{and} \bibinfo{author}{\bibfnamefont{C.}~\bibnamefont{Evoli}},
  \bibinfo{journal}{Phys. Rev.} \textbf{\bibinfo{volume}{D93}},
  \bibinfo{pages}{123007} (\bibinfo{year}{2016}), \eprint{1602.04710}.

\bibitem[{\citenamefont{Aharonian et~al.}(2018)\citenamefont{Aharonian, Peron,
  Yang, Casanova, and Zanin}}]{Aharonian:2018rob}
\bibinfo{author}{\bibfnamefont{F.}~\bibnamefont{Aharonian}},
  \bibinfo{author}{\bibfnamefont{G.}~\bibnamefont{Peron}},
  \bibinfo{author}{\bibfnamefont{R.}~\bibnamefont{Yang}},
  \bibinfo{author}{\bibfnamefont{S.}~\bibnamefont{Casanova}}, \bibnamefont{and}
  \bibinfo{author}{\bibfnamefont{R.}~\bibnamefont{Zanin}}
  (\bibinfo{year}{2018}), \eprint{1811.12118}.

\bibitem[{\citenamefont{Recchia et~al.}(2016)\citenamefont{Recchia, Blasi, and
  Morlino}}]{Recchia:2016}
\bibinfo{author}{\bibfnamefont{S.}~\bibnamefont{Recchia}},
  \bibinfo{author}{\bibfnamefont{P.}~\bibnamefont{Blasi}}, \bibnamefont{and}
  \bibinfo{author}{\bibfnamefont{G.}~\bibnamefont{Morlino}},
  \bibinfo{journal}{Mon. Not. Roy. Astron. Soc.}
  \textbf{\bibinfo{volume}{462}}, \bibinfo{pages}{L88} (\bibinfo{year}{2016}),
  \eprint{1604.07682}.

\bibitem[{\citenamefont{Cerri et~al.}(2017)\citenamefont{Cerri, Gaggero,
  Vittino, Evoli, and Grasso}}]{Cerri:2017joy}
\bibinfo{author}{\bibfnamefont{S.~S.} \bibnamefont{Cerri}},
  \bibinfo{author}{\bibfnamefont{D.}~\bibnamefont{Gaggero}},
  \bibinfo{author}{\bibfnamefont{A.}~\bibnamefont{Vittino}},
  \bibinfo{author}{\bibfnamefont{C.}~\bibnamefont{Evoli}}, \bibnamefont{and}
  \bibinfo{author}{\bibfnamefont{D.}~\bibnamefont{Grasso}},
  \bibinfo{journal}{JCAP} \textbf{\bibinfo{volume}{1710}}, \bibinfo{pages}{019}
  (\bibinfo{year}{2017}), \eprint{1707.07694}.

\bibitem[{\citenamefont{Pothast et~al.}(2018)\citenamefont{Pothast, Gaggero,
  Storm, and Weniger}}]{Pothast:2018}
\bibinfo{author}{\bibfnamefont{M.}~\bibnamefont{Pothast}},
  \bibinfo{author}{\bibfnamefont{D.}~\bibnamefont{Gaggero}},
  \bibinfo{author}{\bibfnamefont{E.}~\bibnamefont{Storm}}, \bibnamefont{and}
  \bibinfo{author}{\bibfnamefont{C.}~\bibnamefont{Weniger}},
  \bibinfo{journal}{JCAP} \textbf{\bibinfo{volume}{1810}}, \bibinfo{pages}{045}
  (\bibinfo{year}{2018}), \eprint{1807.04554}.

\bibitem[{\citenamefont{Bartoli et~al.}(2015)}]{Bartoli:2015}
\bibinfo{author}{\bibfnamefont{B.}~\bibnamefont{Bartoli}} \bibnamefont{et~al.}
  (\bibinfo{collaboration}{ARGO-YBJ}), \bibinfo{journal}{Astrophys. J.}
  \textbf{\bibinfo{volume}{806}}, \bibinfo{pages}{20} (\bibinfo{year}{2015}),
  \eprint{1507.06758}.

\bibitem[{\citenamefont{Abramowski et~al.}(2014)}]{Abramowski:2014}
\bibinfo{author}{\bibfnamefont{A.}~\bibnamefont{Abramowski}}
  \bibnamefont{et~al.} (\bibinfo{collaboration}{H.E.S.S.}),
  \bibinfo{journal}{Phys. Rev.} \textbf{\bibinfo{volume}{D90}},
  \bibinfo{pages}{122007} (\bibinfo{year}{2014}), \eprint{1411.7568}.

\bibitem[{\citenamefont{Abdo et~al.}(2008)}]{Abdo:2008}
\bibinfo{author}{\bibfnamefont{A.~A.} \bibnamefont{Abdo}} \bibnamefont{et~al.},
  \bibinfo{journal}{Astrophys. J.} \textbf{\bibinfo{volume}{688}},
  \bibinfo{pages}{1078} (\bibinfo{year}{2008}), \eprint{0805.0417}.

\bibitem[{\citenamefont{Hao~Zhou and for~the
  HAWC~Collaboration}(2017)}]{HAWCdata}
\bibinfo{author}{\bibfnamefont{C.~D.~R.} \bibnamefont{Hao~Zhou}}
  \bibnamefont{and} \bibinfo{author}{\bibfnamefont{G.~V.} \bibnamefont{for~the
  HAWC~Collaboration}}, \bibinfo{journal}{talk at ICRC}
  (\bibinfo{year}{2017}).

\bibitem[{\citenamefont{Pagliaroli et~al.}(2016)\citenamefont{Pagliaroli,
  Evoli, and Villante}}]{Pagliaroli:2016}
\bibinfo{author}{\bibfnamefont{G.}~\bibnamefont{Pagliaroli}},
  \bibinfo{author}{\bibfnamefont{C.}~\bibnamefont{Evoli}}, \bibnamefont{and}
  \bibinfo{author}{\bibfnamefont{F.~L.} \bibnamefont{Villante}},
  \bibinfo{journal}{JCAP} \textbf{\bibinfo{volume}{1611}}, \bibinfo{pages}{004}
  (\bibinfo{year}{2016}), \eprint{1606.04489}.

\bibitem[{\citenamefont{Villante et~al.}(2017)\citenamefont{Villante, Evoli,
  and Pagliaroli}}]{Villante:2017}
\bibinfo{author}{\bibfnamefont{F.~L.} \bibnamefont{Villante}},
  \bibinfo{author}{\bibfnamefont{C.}~\bibnamefont{Evoli}}, \bibnamefont{and}
  \bibinfo{author}{\bibfnamefont{G.}~\bibnamefont{Pagliaroli}},
  \bibinfo{journal}{PoS} \textbf{\bibinfo{volume}{NOW2016}},
  \bibinfo{pages}{052} (\bibinfo{year}{2017}).

\bibitem[{\citenamefont{Pagliaroli and Villante}(2018)}]{Pagliaroli:2017}
\bibinfo{author}{\bibfnamefont{G.}~\bibnamefont{Pagliaroli}} \bibnamefont{and}
  \bibinfo{author}{\bibfnamefont{F.~L.} \bibnamefont{Villante}},
  \bibinfo{journal}{JCAP} \textbf{\bibinfo{volume}{1808}}, \bibinfo{pages}{035}
  (\bibinfo{year}{2018}), \eprint{1710.01040}.

\bibitem[{\citenamefont{Lipari and Vernetto}(2018)}]{Lipari:2018}
\bibinfo{author}{\bibfnamefont{P.}~\bibnamefont{Lipari}} \bibnamefont{and}
  \bibinfo{author}{\bibfnamefont{S.}~\bibnamefont{Vernetto}},
  \bibinfo{journal}{Phys. Rev.} \textbf{\bibinfo{volume}{D98}},
  \bibinfo{pages}{043003} (\bibinfo{year}{2018}), \eprint{1804.10116}.

\bibitem[{\citenamefont{Abdalla et~al.}(2018)}]{HESS:2018}
\bibinfo{author}{\bibfnamefont{H.}~\bibnamefont{Abdalla}} \bibnamefont{et~al.}
  (\bibinfo{collaboration}{HESS}), \bibinfo{journal}{Astron. Astrophys.}
  \textbf{\bibinfo{volume}{612}}, \bibinfo{pages}{A1} (\bibinfo{year}{2018}),
  \eprint{1804.02432}.

\bibitem[{\citenamefont{Dembinski et~al.}(2018)\citenamefont{Dembinski, Engel,
  Fedynitch, Gaisser, Riehn, and Stanev}}]{Dembinski:2017}
\bibinfo{author}{\bibfnamefont{H.~P.} \bibnamefont{Dembinski}},
  \bibinfo{author}{\bibfnamefont{R.}~\bibnamefont{Engel}},
  \bibinfo{author}{\bibfnamefont{A.}~\bibnamefont{Fedynitch}},
  \bibinfo{author}{\bibfnamefont{T.}~\bibnamefont{Gaisser}},
  \bibinfo{author}{\bibfnamefont{F.}~\bibnamefont{Riehn}}, \bibnamefont{and}
  \bibinfo{author}{\bibfnamefont{T.}~\bibnamefont{Stanev}},
  \bibinfo{journal}{PoS} \textbf{\bibinfo{volume}{ICRC2017}},
  \bibinfo{pages}{533} (\bibinfo{year}{2018}), \bibinfo{note}{[35,533(2017)]},
  \eprint{1711.11432}.

\bibitem[{\citenamefont{Ahlers et~al.}(2016)\citenamefont{Ahlers, Bai, Barger,
  and Lu}}]{Ahlers:2015}
\bibinfo{author}{\bibfnamefont{M.}~\bibnamefont{Ahlers}},
  \bibinfo{author}{\bibfnamefont{Y.}~\bibnamefont{Bai}},
  \bibinfo{author}{\bibfnamefont{V.}~\bibnamefont{Barger}}, \bibnamefont{and}
  \bibinfo{author}{\bibfnamefont{R.}~\bibnamefont{Lu}}, \bibinfo{journal}{Phys.
  Rev.} \textbf{\bibinfo{volume}{D93}}, \bibinfo{pages}{013009}
  (\bibinfo{year}{2016}), \eprint{1505.03156}.

\bibitem[{\citenamefont{Green}(2015)}]{Green:2015}
\bibinfo{author}{\bibfnamefont{D.~A.} \bibnamefont{Green}},
  \bibinfo{journal}{Mon. Not. Roy. Astron. Soc.}
  \textbf{\bibinfo{volume}{454}}, \bibinfo{pages}{1517} (\bibinfo{year}{2015}),
  \eprint{1508.02931}.

\bibitem[{Gal()}]{Galprop}
\urlprefix\url{http://galprop.stanford.edu/}.

\bibitem[{\citenamefont{Gaggero
  et~al.}(2015{\natexlab{b}})\citenamefont{Gaggero, Urbano, Valli, and
  Ullio}}]{Gaggero:2014}
\bibinfo{author}{\bibfnamefont{D.}~\bibnamefont{Gaggero}},
  \bibinfo{author}{\bibfnamefont{A.}~\bibnamefont{Urbano}},
  \bibinfo{author}{\bibfnamefont{M.}~\bibnamefont{Valli}}, \bibnamefont{and}
  \bibinfo{author}{\bibfnamefont{P.}~\bibnamefont{Ullio}},
  \bibinfo{journal}{Phys. Rev.} \textbf{\bibinfo{volume}{D91}},
  \bibinfo{pages}{083012} (\bibinfo{year}{2015}{\natexlab{b}}),
  \eprint{1411.7623}.

\bibitem[{\citenamefont{Gaggero et~al.}(2017)\citenamefont{Gaggero, Grasso,
  Marinelli, Taoso, and Urbano}}]{Gaggero:2017jts}
\bibinfo{author}{\bibfnamefont{D.}~\bibnamefont{Gaggero}},
  \bibinfo{author}{\bibfnamefont{D.}~\bibnamefont{Grasso}},
  \bibinfo{author}{\bibfnamefont{A.}~\bibnamefont{Marinelli}},
  \bibinfo{author}{\bibfnamefont{M.}~\bibnamefont{Taoso}}, \bibnamefont{and}
  \bibinfo{author}{\bibfnamefont{A.}~\bibnamefont{Urbano}},
  \bibinfo{journal}{Phys. Rev. Lett.} \textbf{\bibinfo{volume}{119}},
  \bibinfo{pages}{031101} (\bibinfo{year}{2017}), \eprint{1702.01124}.

\bibitem[{\citenamefont{Kappes et~al.}(2007)\citenamefont{Kappes, Hinton,
  Stegmann, and Aharonian}}]{Kappes:2006}
\bibinfo{author}{\bibfnamefont{A.}~\bibnamefont{Kappes}},
  \bibinfo{author}{\bibfnamefont{J.}~\bibnamefont{Hinton}},
  \bibinfo{author}{\bibfnamefont{C.}~\bibnamefont{Stegmann}}, \bibnamefont{and}
  \bibinfo{author}{\bibfnamefont{F.~A.} \bibnamefont{Aharonian}},
  \bibinfo{journal}{Astrophys. J.} \textbf{\bibinfo{volume}{656}},
  \bibinfo{pages}{870} (\bibinfo{year}{2007}), \bibinfo{note}{[Erratum:
  Astrophys. J.661,1348(2007)]}, \eprint{astro-ph/0607286}.

\bibitem[{\citenamefont{Egberts}(2018)}]{Egberts:2017apw}
\bibinfo{author}{\bibfnamefont{K.}~\bibnamefont{Egberts}},
  \bibinfo{journal}{PoS} \textbf{\bibinfo{volume}{ICRC2017}},
  \bibinfo{pages}{684} (\bibinfo{year}{2018}).

\end{thebibliography}
   
\clearpage

\appendix

\section{The contribution to the total gamma-ray flux due to HESS unresolved sources}
\label{App:UnresolvedSources}

The HESS source catalogue \cite{HESS:2018} can be considered complete for sources producing an integrated flux at Earth $\Phi(E_\gamma \ge 1 \,{\rm TeV})$ larger than $\Phi_{\rm th} = 10\% \; \Phi_{\rm CRAB}$, where $\Phi_{\rm CRAB}=  2.26\cdot 10^{-11} \,{\rm cm}^{-2}\,{\rm s}^{-1}$ is the flux emitted by CRAB nebula above 1 TeV.
If we take the CRAB emission spectrum as a reference, a source at a distance $D$ with intrinsic luminosity $L$ (above 1 TeV) produces a flux at Earth given by $\Phi = \Phi_{\rm CRAB}\, (L/L_{\rm CRAB})\, (D/D_{\rm CRAB})^{-2}$, where $L_{\rm CRAB}=5\cdot 10^{34}\,{\rm erg}\,{\rm s}^{-1}$ and $D_{\rm CRAB}= 2\,{\rm kpc}$ are CRAB luminosity and distance, respectively.
The produced flux is larger than $\Phi_{\rm th}$ only if the source distance is $D \le \overline{D}(L)$ where:
\begin{equation}
\overline{D}(L) \equiv 6.3 \,{\rm kpc} \sqrt{L/L_{\rm CRAB}}
\label{Eq:horizon}
\end{equation}
The above estimate shows that a relatively small region of the Galaxy is fully resolved by HESS, even if sources are assumed to be very luminous.

In the assumption of fixed source intrinsic luminosity $L$, the cumulative contribution of sources to the gamma-ray flux for $E_\gamma\ge 1\,\rm{TeV}$ observed from the direction ${\hat n}$ can be calculated as:
\begin{equation}
\varphi_{\gamma,\rm S}({\hat n}, L) = \Phi_{\rm th} \, \overline{D}(L)^2 \int_{0}^{\infty} dl  \, f_{\rm S}({\bf r}_\odot + {\hat n} l)
\end{equation}
where $f_{\rm S}({\bf r})$ represents the source distribution in the Galaxy. The total flux can be divided in two contributions, "fully resolved" and "partially resolved" sources. The contribution of fully resolved sources with $\Phi\ge \Phi_{\rm th}$ can be calculated as:
\begin{equation}
\varphi^{(fr)}_{\gamma,\rm S}({\hat n}, L) = \Phi_{\rm th} \, \overline{D}(L)^2 \int_{0}^{\overline{D}(L)} dl  \, f_{\rm S}({\bf r}_\odot + {\hat n} l)
\end{equation}
Following this prescription, the flux from sources with $\Phi\le\Phi_{\rm th}$, that maybe are not resolved being very faint and/or extended, is instead given by: 
\begin{equation}
\varphi^{(pr)}_{\gamma,\rm S}({\hat n},L) = \Phi_{\rm th} \, \overline{D}(L)^2 \int_{\overline{D}(L)}^{\infty} dl  \, f_{\rm S}({\bf r}_\odot + {\hat n} l).
\end{equation}
Since the cutoff distance $\overline{D}(L)$ can be relatively small with respect to the galactic scales, we expect that the ratio $\varphi^{(pr)}/\varphi^{(fr)}$ may be large, in particular when looking at small galactic latitudes where the source density along the line-of-sight is not expected to vanish for large distances.
In order to estimate the relevance of "partially resolved" contribution, we calculate the ratio:
\begin{equation}
R(L) = \frac{\Phi^{(pr)}_{\rm S}(L)}{\Phi^{(fr)}_{\rm S}(L)}
\end{equation}
where the fluxes:
\begin{eqnarray}
\nonumber
\Phi^{(pr)}_{\rm S}(L)&\equiv& \int_{\Delta \Omega} d\Omega\;  \varphi^{(pr)}_{\gamma,\rm S}({\hat n},L)\\
\Phi^{(fr)}_{\rm S}(L)&\equiv& \int_{\Delta \Omega} d\Omega\;  \varphi^{(fr)}_{\gamma,\rm S}({\hat n},L)\\
\nonumber
\end{eqnarray}
give the "partially resolved" and "fully resolved" source emission integrated over the angular region ${\Delta \Omega}$ defined as $-60^\circ \le l \le 60^\circ$ and $-3^\circ \le b \le 3^\circ$. 
If we take $L=L_{\rm CRAB}$ and we assume that the source distribution $f_{\rm S}({\bf r})$ is proportional to the SNR number density given in \cite{Green:2015}, we obtain $R=0.9$. By considering a uniform cylindrical source distribution with radius $R=15\,{\rm kpc}$ and thickness $H=0.2\,{\rm kpc}$, we obtain $R=0.8$.

The above results can be used to estimate the contribution to total gamma-ray flux by sources unresolved by HESS. In the angular region $\Delta \Omega$, the HESS experiment observes 29 sources with fluxes larger than $\Phi_{\rm th}=10\%\,\Phi_{\rm CRAB}$, corresponding to a "fully resolved" flux $\Phi^{(fr)}_{\rm S} = 7.0 \, \Phi_{\rm CRAB}$. This represents a fraction $\omega = 0.84$ of the cumulative resolved flux $\Phi^{(r)}_{\rm S}$ that  includes 42 additional sources with $\Phi \le \Phi_{\rm th}$. 
Taking this into account and considering that the total flux $\Phi_{\gamma,\rm S}$ is given by the sum of fully resolved and partially resolved contribution, we obtain:
\begin{equation}
\Phi_{\gamma,\rm S} = \Phi^{(r)}_{\gamma,\rm S}(1+\eta) 
\end{equation}
where the parameter $(1+\eta) \equiv \omega + \omega R$ is defined as in Eq.\ref{eta} and can be relatively large. If we take $L=L_{\rm CRAB}$ we obtain $\eta \simeq 0.6$, i.e. unresolved sources account for an additional contribution equal to $\sim 60\%$ of the resolved signal.
Despite the large uncertainty, the above estimate allows us to conclude that resolved and unresolved sources  plausibly give comparable contributions to the total gamma-ray observed signal, in substantial agreement with results presented by \cite{Egberts:2017apw}.

\end{document}